\documentstyle[12pt]{article}
\newcommand{\dmsq}{$\Delta m^2$}
\newcommand{\mixing}{$\sin^2 2\theta$}

\newcommand {\omutau} {$\nu_\mu \rightarrow \nu_\tau$}
\newcommand {\omuste} {$\nu_\mu \rightarrow \nu_{\mathrm{s}}$}
\newcommand {\omue} {$\nu_\mu \rightarrow \nu_e$}

\newcommand {\xedec} {$\tau^- \rightarrow e^-\nu\bar{\nu}$}
\newcommand {\xmudec} {$\tau^- \rightarrow \mu^-\nu\bar{\nu}$}

\def\xa1dec{$\tau^- \rightarrow a_1^-\nu$}
\def\za1dec{$\tau^- \rightarrow \pi^- \pi^+ \pi^- \nu$}
\def\za1op{$\tau^- \rightarrow \pi^- \pi^0 \pi^0 \nu$}
\def\yrho_2pi{$\rho^- \rightarrow \pi^-\pi^0$}
\def\ya1_3pi{$a_1^- \rightarrow \pi^-\pi^+\pi^-$}
\def\ypi_gg{$\pi^0 \rightarrow \gamma\gamma$}

\begin {document}
\title
{A method for detecting $\nu_\tau$ appearance in the spectra of quasielastic 
CC events}
\author
{A.E. Asratyan\thanks{Corresponding author. Tel.: 095-237-0079. E-mail
address: asratyan@vxitep.itep.ru.},
G.V. Davidenko, A.G. Dolgolenko, V.S. Kaftanov,\\
M.A. Kubantsev\thanks{Now at Fermi National Accelerator Laboratory, Batavia, 
IL 60510, USA.},
and V.S. Verebryusov\\
\normalsize{\it Institute of Theoretical and Experimental Physics,}
\normalsize{\it Moscow 117259, Russia}}
\date {\today}
\maketitle

\begin{abstract}
A method for detecting the transition \omutau\ in long-baseline accelerator
experiments, that consists in comparing the far-to-near ratios of the spectra
of quasielastic CC events generated by  high- and low-energy beams of muon 
neutrinos, is proposed. The test may be accessible to big water Cherenkov
detectors and iron--scintillator calorimeters, and is limited by statistics 
rather than systematics.
\end{abstract}
\clearpage

     The data of Super-Kamiokande \cite{sk_on_omutau} favor the transition
\omutau\ as the source of the deficit of muon neutrinos from the atmosphere.
However, this still has to be verified by directly observing $\nu_\tau$
appearance in accelerator long-baseline experiments. The options discussed
thus far, all involving fine instrumentation on a large scale, are to detect 
the secondary $\tau$ by range in emulsion \cite{opera}, by Cherenkov light 
of the $\tau$ \cite{cherenkov}, or by the transverse momentum carried away 
by the decay neutrino(s) \cite{icarus}. By contrast, in this paper we wish 
to formulate a $\tau$ signature that is solely based on the energy spectra 
of CC events, and therefore should be accessible to water Cherenkov 
detectors and to relatively coarse calorimeters with
muon spectrometry. We assume that the experiment includes a near detector of 
the same structure as the far detector, irradiated by the same neutrino beam 
but over a short baseline that rules out any significant effects of neutrino
oscillations \cite{numi}. Thereby, the systematic uncertainties in comparing
the interactions of primary and oscillated neutrinos are largely eliminated.
Our aim is to distinguish the muonic decays of $\tau$ leptons against the 
background of $\nu_\mu$-induced CC events. In order to minimize the effects 
of $\nu_\mu$ disappearance, the data collected with a harder $\tau$-producing
beam are compared with those for a softer reference beam in which $\tau$ 
production is suppressed by the threshold effect. The analysis is restricted 
to quasielastics (QE), that is, to neutrino events featuring a muon and small 
hadronic energy.

     As soon as the first maximum of the oscillation lies below the mean
energy of muon neutrinos in the beam, or
$\Delta m^2 L / \langle E_\nu \rangle < 1.24$ eV$^2$km/GeV,
much of the signal from QE production and muonic decay of the $\tau$ is
at relatively low values of visible energy $E$. That is because the tau 
neutrinos arising from \omutau\ are softer on average than muon neutrinos, 
the threshold effect is relatively mild for quasielastics, and a large 
fraction of incident energy is taken away by the two neutrinos from \xmudec. 
Let $f(E)$ be the spectrum of QE events observed in the far detector,
$n(E)$---the spectrum of similar events in the near detector that has been 
extrapolated and normalized to the location of the far detector, and 
$R(E)$---the ratio of the two: $R(E) = f(E) / n(E)$. In the case of $\nu_\mu$
disappearance through the transitions \omue\ or \omuste, where 
$\nu_{\mathrm{s}}$ is the hypothesized sterile neutrino, the ratio $R$ for 
the harder beam should be identically equal to that for the softer beam: 
$R_{\mathrm{hard}}(E) = R_{\mathrm{soft}}(E)$. However, in the case of 
\omutau\ this equation is violated by the process of $\tau$ production and 
muonic decay, that predominantly occurs in the harder beam and shows up as a 
low-$E$ enhancement of the corresponding "far" spectrum 
$f_{\mathrm{hard}}(E)$. This causes the ratio $R_{\mathrm{hard}}$ to exceed 
$R_{\mathrm{soft}}$ towards low values of visible energy $E$. The latter 
effect, that may provide a specific signature of $\nu_\tau$ appearance, is 
investigated in this paper.

     In the simulations reported below, the harder (or $\tau$-producing) and
softer (or reference) beams of muon neutrinos are respectively assigned as
the high- and low-energy beams from the Main Injector (MI) at Fermilab, as 
foreseen by the NuMI program \cite{numi}. The mean $\nu_\mu$ energies in
these beams are close to 12 and 5 GeV, respectively. Again as in the NuMI 
program, a baseline of 730 km is assumed throughout. The systematic 
uncertaities on the near spectra $n(E)$ are neglected\footnote{The
systematic uncertainty on $n(E)$, that arises from correcting the spectrum
in the near detector for the neutrino source not being pointlike, is analyzed 
in \cite{systematics}. For the statistics considered in this paper, the 
overall uncertainty on the effect is still dominated by statistical errors.
Furthermore, we may expect that systematic uncertainties on $n(E)$ for the 
high- and low-energy beams from the Main Injector are correlated and,
therefore, will partially cancel in the difference of far-to-near ratios for 
these two beams.}.
The two detector types considered are an iron--scintillator calorimeter and 
a water Cherenkov detector. 

     Charged-current interactions of the $\nu_\mu$ and $\nu_\tau$ are 
generated using the NEUGEN package that is based on the Soudan-2 Monte 
Carlo \cite{generator}. This generator takes full account of exclusive
channels like quasielastics proper and excitation of baryon resonances, 
that are important for our analysis of CC events with small hadronic
energy.

     The iron--scintillator calorimeter is assumed to be the MINOS 
detector \cite{numi} that is in construction stage. The detector response 
is not simulated in full detail; instead, the resolution in muon energy is 
approximated as $\delta E_\mu = 0.11 \times E_\mu^{\mathrm{true}}$ and in 
energy transfer to hadrons---as $\delta\nu$ (GeV) = 
$0.55 \times \sqrt{\nu^{\mathrm{true}}}$ \cite{resolutions}. Quasielastic 
events are selected as those with $E_\mu > 800$ MeV and $\nu < 1$ GeV, where 
$E_\mu$ and $\nu$ are the smeared values of muon energy and of energy 
transfer to hadrons, respectively\footnote{These selections should be 
viewed as illustrative. The actual selections will be based on a detailed
simulation of detector response to CC events with small hadronic energy.}.
Given the characteristic topology of such events in the detector (a single 
track traversing more than three nuclear interaction lengths in iron plus a
few scintillator hits near the primary vertex), we assume that they are 
reconstructed with 100\% efficiency and that the background from pion
punchthrough is insignificant. The visible energy of a CC event, $E$, is
estimated in terms of smeared quantities: $E = E_\mu + \nu$. 

     We also analyze the performance of a water Cherenkov detector,
in which $\nu_\mu$-induced quasielastics have a characteristic signature of
muonlike single-ring events \cite{sk_muons}. Analyzing the multi-GeV exiting
muons, particularly in the smaller near detector, may require a device of
the AQUA-RICH type \cite{aquarich} or an instrumented muon absorber 
downstream of the water tank. In our simulation for the water Cherenkov 
detector, QE events are selected as those featuring a single muon with 
momentum above 200 MeV, no additional charged secondaries with momenta above 
the Cherenkov threshold in water, and no $\pi^0$ mesons in the final state. 
For the quasielastic reaction $\nu_\mu n \rightarrow \mu^-p$, this implies 
an energy transfer to the nucleon of less than 0.47 GeV. (Some 70\% of thus 
selected events are due to quasielastics proper.) Muon energy is then smeared
according to $\delta E_\mu = 0.05 \times E_\mu^{\mathrm{true}}$, and neutrino 
energy is estimated assuming quasielastic scattering on a free neutron:
\begin{equation}
E = \frac {m_N E_\mu - m_\mu^2/2} {m_N - E_\mu + p_\mu \cos \theta_\mu} .
\end{equation}
Here, $m_N$ and $m_\mu$ are the neutron and muon masses, and $p_\mu$ and
$\theta_\mu$ are the muon momentum and angle relative to incident neutrino, 
respectively.

\begin{figure}[p]
\vspace{18 cm}
\includegraphics{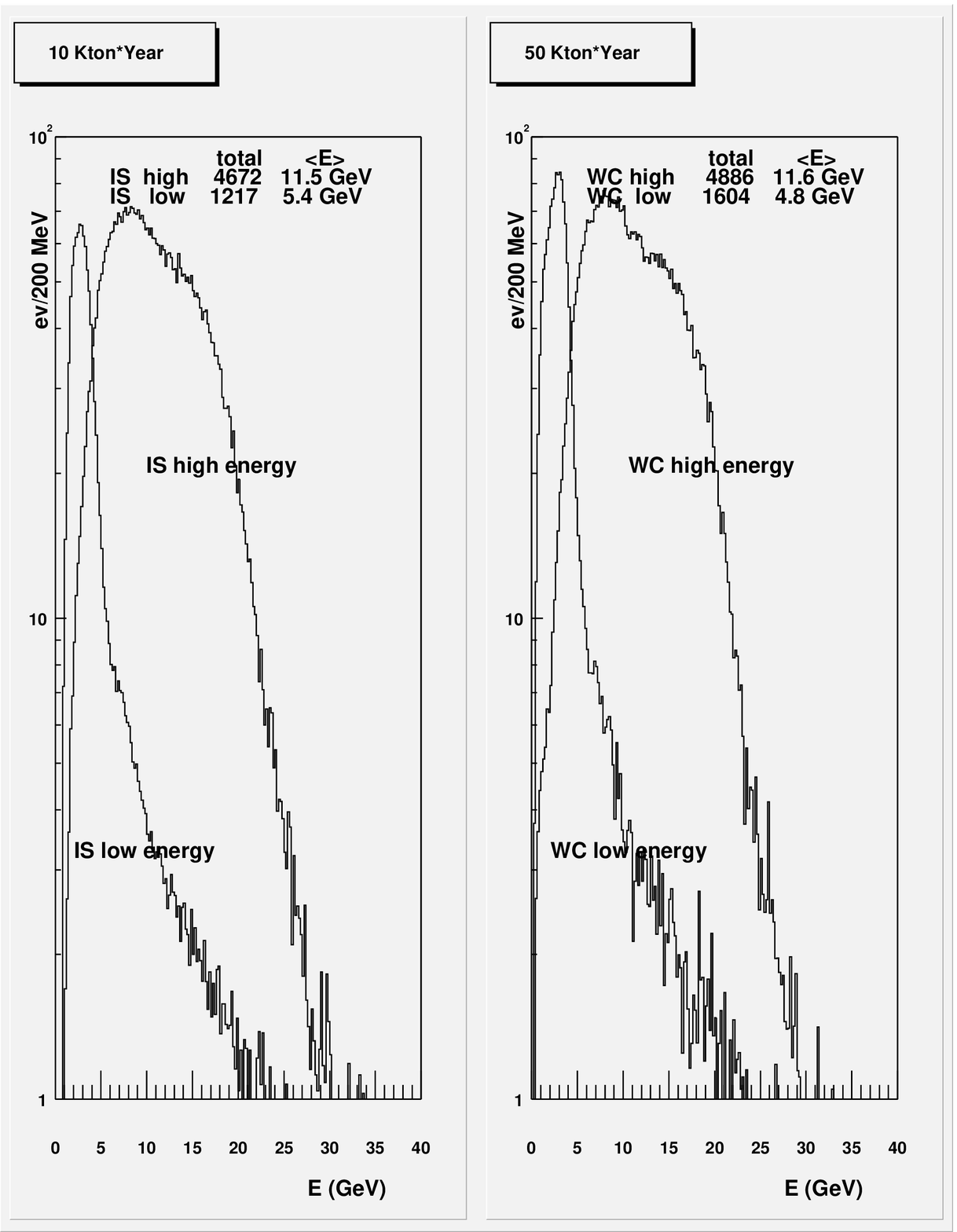}
\caption
{The oscillation-free "near" spectra of $\nu_\nu$-induced quasielastic 
events, $n(E)$, for the iron--scintillator (on the left) and water 
Cherenkov (on the right)
detectors irradiated by the low-energy and high-energy beams from the Main 
Injector. The assumed exposures are 10 and 50 kton--years, respectively.}
\label{spectra}
\end{figure}

\begin{figure}
\vspace{18 cm}
\includegraphics{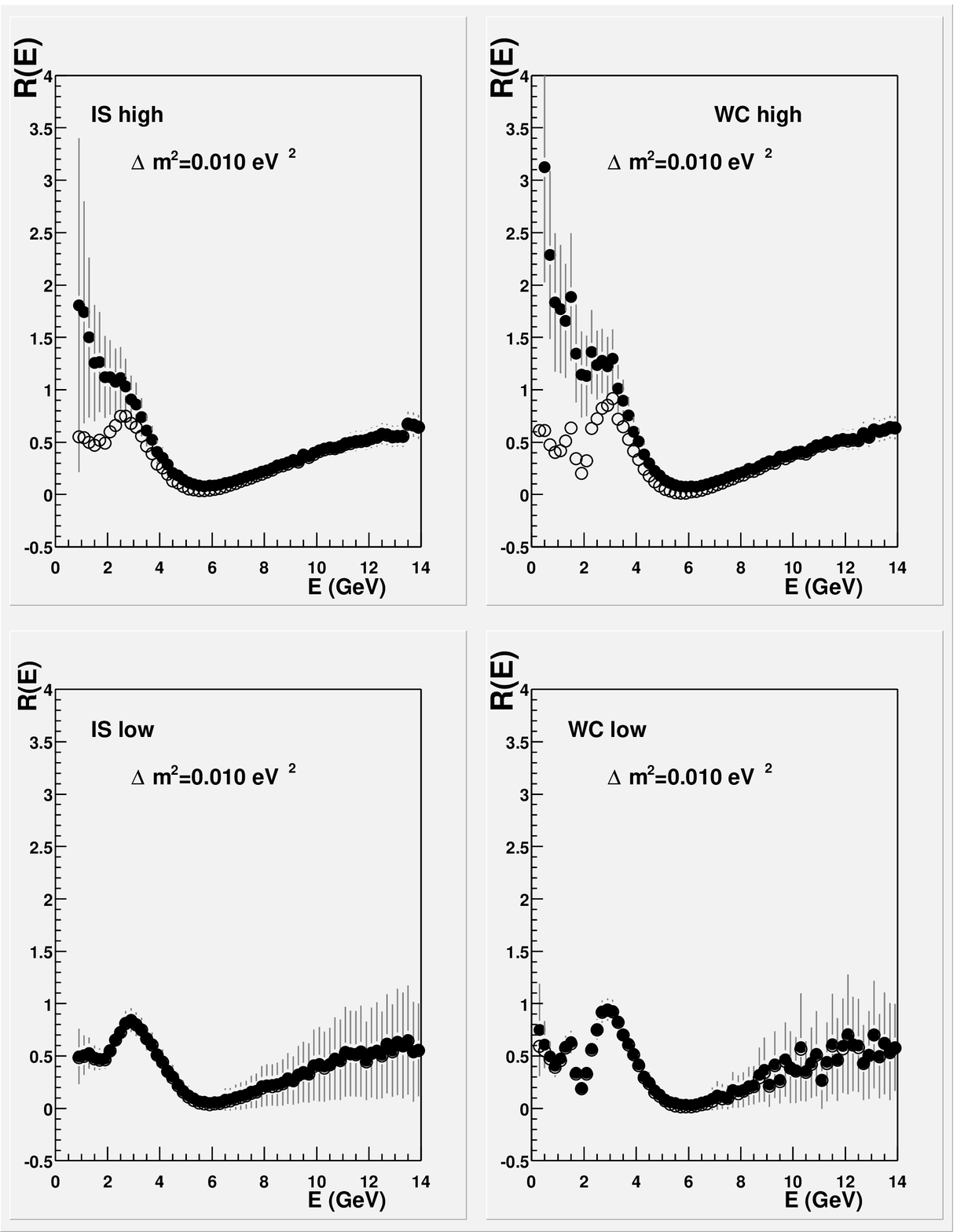}
\caption
{Assuming the transitions \omutau\ and \omuste\ (solid and open dots, 
respectively) driven by \dmsq\ = 10$^{-2}$ eV$^2$ and \mixing\ = 1, 
the far-to-near ratio $R(E) = f(E) / n(E)$ for quasielastic events produced 
by the MI high- and low-energy beams (top and bottom panels, respectively)
in the iron--scintillator and water Cherenkov detectors (left- and right-hand
panels, respectively). The error bars on $R(E)$ for the transition \omutau\ 
are the statistical uncertainties corresponding to 10 (50) kton--year 
exposures of the iron--scintillator (water Cherenkov) detector 
in either beam.}
\label{ratios}
\end{figure}

\begin{figure}
\vspace{18 cm}
\includegraphics{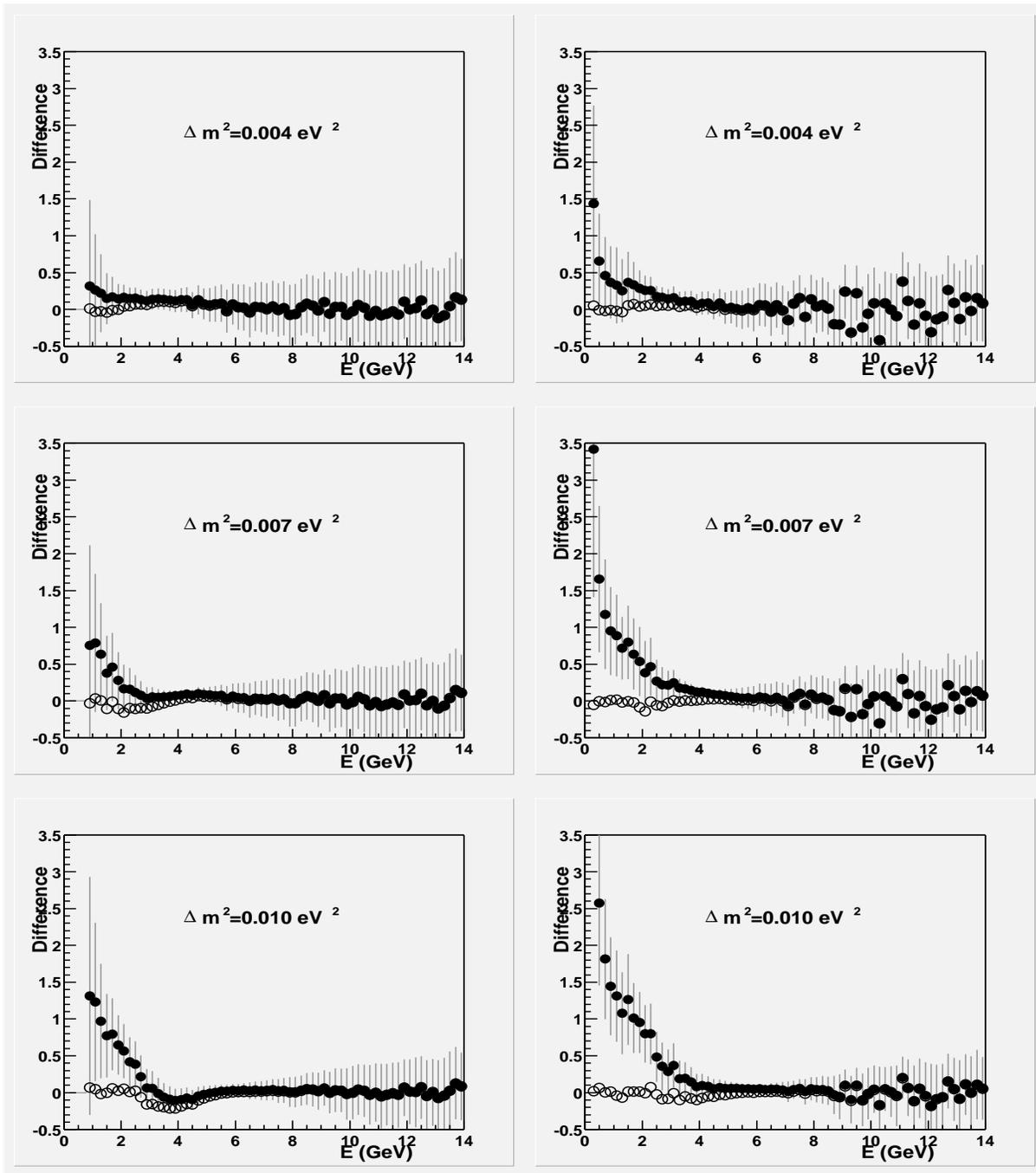}
\caption
{Assuming either \omutau\ or \omuste\ (solid and open dots, respectively)
with maximal mixing, the difference $\Delta R(E)$ between the far-to-near 
ratios for the MI high- and low-energy beams. Simulated data for the
iron--scintillator and water Cherenkov detectors are shown in the left-
and right-hand panels, respectively. The top, middle, and bottom panels are 
for \dmsq\ = 0.004, 0.007, and 0.010 eV$^2$, respectively. Depicted by 
error bars are the statistical errors that correspond to 10 (50) kton--year 
exposures of the iron--scintillator (water Cherenkov) detector 
in either beam.}
\label{difference}
\end{figure}

     Shown in Fig. \ref{spectra} are the oscillation-free near spectra of 
selected QE events, $n(E)$, for the two beams and two detectors considered.
In the absence of oscillations, 10 (50) kton--year exposures of the 
iron--scintillator (water Cherenkov) detector in the softer and harder beams 
will yield some 1200 (1600) and 4700 (4900) $\nu_\mu$-induced QE events, 
respectively\footnote{In all 
numerical estimates, we do not take into account the discussed upgrade of 
the proton driver at Fermilab \cite{driver} that may result in a substantial 
increase of neutrino flux from the Main Injector.}. Assuming either \omutau\ 
or \omuste\ driven by \dmsq\ = 0.01 eV$^2$ and maximal mixing of 
\mixing\ = 1, the far-to-near ratios $R(E)$ for either beam and either 
detector are illustrated in Fig. \ref{ratios}\footnote{In the Figures, 
statistical fluctuations are suppressed for the data points themselves, but 
the error bars are for the statistics as indicated.}.
That the ratios $R(E)$ for the transitions \omutau\ and \omuste\ diverge 
towards low values of $E$ is evident for the harder beam in which $\tau$ 
production is not suppressed. Again considering \omutau\ and \omuste\ with 
maximal mixings, in Fig. \ref{difference} we plot the difference
\begin{displaymath}
\Delta R(E) = R_{\mathrm{hard}} - R_{\mathrm{soft}}
\end{displaymath}
for various values of \dmsq. Indeed, at visible energies below some 4 GeV
$\Delta R(E)$ deviates from zero for the transition \omutau, while staying
close to zero for \omuste. This deviation may be viewed as a signature of
$\nu_\tau$ appearance. The naive expectation for \omuste, $\Delta R(E) = 0$, 
is violated by the smearing of neutrino energy, and therefore is better
fulfilled for the water Cherenkov detector.

\begin{figure}
\vspace{18 cm}
\includegraphics{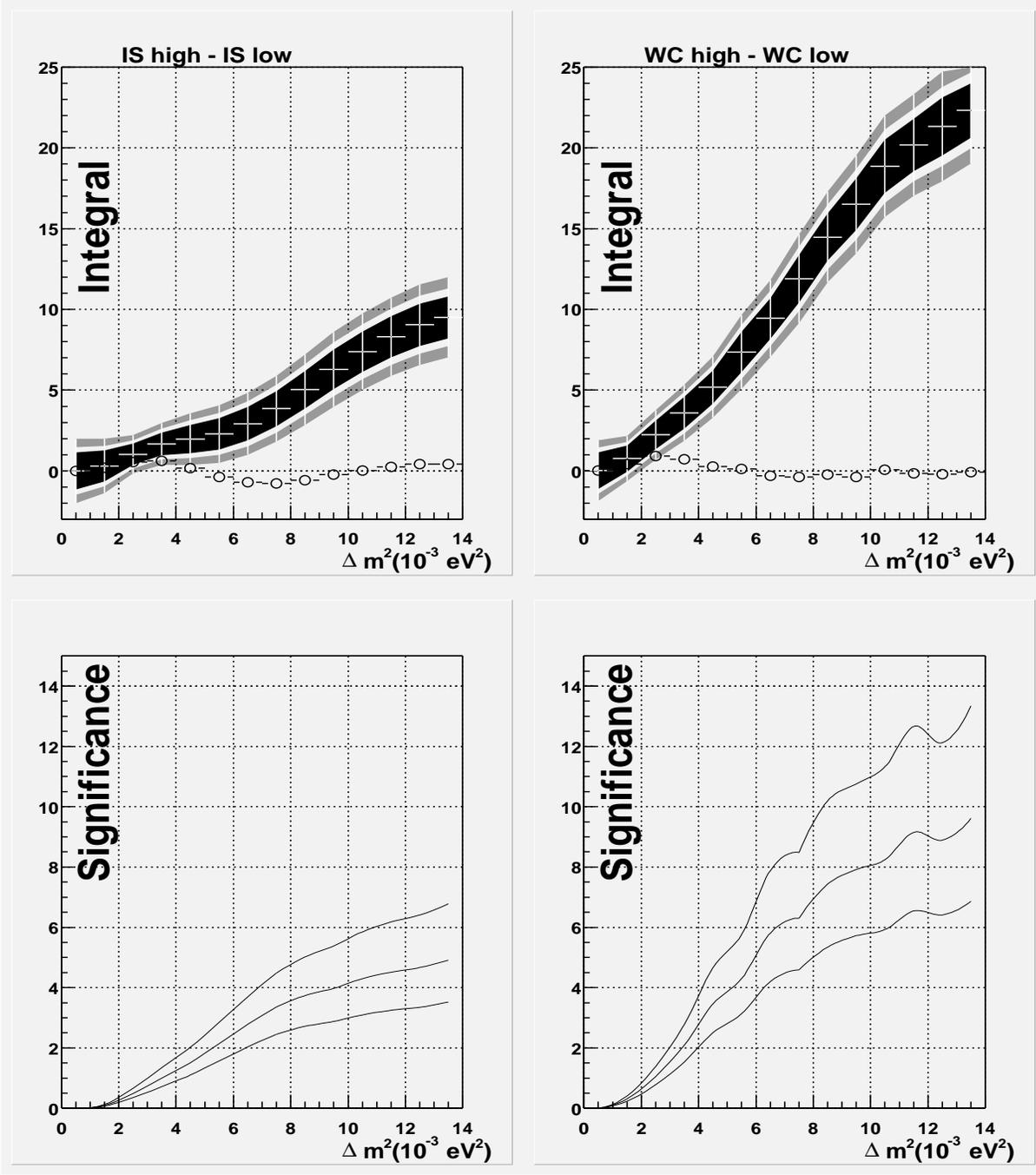}
\caption
{Assuming either \omutau\ or \omuste\ (solid and open dots, respectively)
with maximal mixing, the integrated difference $S$ (see text) as a function 
of \dmsq\ for the iron--scintillator (top left) and water Cherenkov (top
right) detectors. Shown by successive error corridors are the statistical 
errors on $S$(\dmsq) corresponding to 10(50), 20(100), and 40(200) 
kton--year exposures of the iron--scintillator (water Cherenkov) detector
in the $\tau$-producing beam. The bottom panels show the statistical 
significance of the $\tau$ signal for either case.}
\label{integral}
\end{figure}

     By the time the proposed test can be implemented, the actual value of
\dmsq\ will probably be estimated to some 10\% by analyzing the $\nu_\mu$
disappearance in the MI low-energy beam \cite{resolutions}. 
Given the value of \dmsq, a consistent approach would be to fit $\Delta R(E)$
to the predicted shape in order to estimate the mixing between the muon and 
tau neutrinos. A cruder measure of the effect is provided by the integral 
$S = \int \Delta R(E)dE$, which we estimate between $E = 1$ and 3 GeV for the
iron--scintillator detector, and between $E = 0.5$ and 3 GeV for the 
water Cherenkov detector. Allowing for either \omutau\ and \omuste\ with 
maximal mixing, the respective integrals $S$(\omutau) and $S$(\omuste) are 
plotted in Fig. \ref{integral} as functions of \dmsq. 

     Since the reference beam produces many more low-energy events than the 
harder beam, the statistical error on the integral $S$ is largely determined 
by the statistics accumulated with the latter. Therefore, we fix the exposure
of the iron--scintillator (water Cherenkov) detector in the reference beam at
10 (50) kton--years, and assume a similar or bigger exposure in the harder 
beam: 10(50), 20(100), or 40(200) kton--years. The respective statistical 
uncertainties on $S$(\omutau) are illustrated by successive error corridors 
in Fig. \ref{integral}. And finally, dividing the difference between 
$S$(\omutau) and $S$(\omuste) by the statistical error on $S$(\omutau), we 
estimate the statistical significance of the enhancement that is also 
depicted in Fig. \ref{integral}. 

     We estimate that at a level of 3$\sigma$, the 10(50), 20(100), and 
40(200) kton--year exposures of the iron--scintillator (water Cherenkov) 
detector in the MI high-energy beam will allow to probe $\nu_\tau$ appearance
down to the \dmsq\ values of some 0.008 (0.005), 0.006 (0.004), and 
0.005 (0.003) eV$^2$, respectively. Thus in the NuMI program with the 5.4-kton 
MINOS detector and with the existing Proton Booster \cite{numi}, the proposed 
test may be sensitive to \dmsq\ values in the Kamiokande-allowed region 
\cite{kamioka}, but not below some $5\times 10^{-3}$ eV$^2$ as suggested by 
the more recent results of Super-Kamiokande \cite{superkamioka}. On the 
other hand, a big ($\sim 100$ kton) water Cherenkov detector would allow
to probe the transition \omutau\ over a large portion of the \dmsq\ region 
favored by the analysis of atmospheric neutrinos in 
Super-Kamiokande\footnote{Apart from muonic decays of QE-produced $\tau$
leptons, a water Cherenkov detector may also allow to detect the electronic 
decays \xedec\ in the spectra of single-ring electronlike events. This will 
require good understanding of the contribution of neutral-current $\pi^0$ 
production to electronlike events (see, for example, \cite{jhf}).}.

     To conclude, we have proposed a test of $\nu_\tau$ appearance that 
consists in comparing the far-to-near ratios of the spectra of quasielastic 
CC events generated by different beams of muon neutrinos, and therefore may 
be accessible to water Cherenkov detectors and to calorimeters with muon 
spectrometry. The test is limited by statistics rather than systematics, and 
its significance crucially depends on the exposure in the harder beam in 
particular.

\end{document}